\documentclass[fleqn,10pt]{wlscirep}
\title{Topological states and phase transitions in Sb$_2$Te$_3$-GeTe multilayers}

\author[1]{Thuy-Anh Nguyen}
\author[1]{Dirk Backes}
\author[1]{Angadjit Singh}
\author[1]{Rhodri Mansell}
\author[1]{Crispin Barnes}
\author[1]{David A. Ritchie}
\author[2]{Gregor Mussler}
\author[2]{Martin Lanius}
\author[2]{Detlev Gr\"utzmacher}
\author[1,*]{Vijay Narayan}
\affil[1]{Cavendish Laboratory, Department of Physics, University of Cambridge, J. J. Thomson Avenue, Cambridge CB3 0HE, United Kingdom}
\affil[2]{Peter Gr\"unberg Institute (PGI-9), Forschungszentrum J\"ulich, 52425 J\"ulich, Germany}

\affil[*]{vn237@cam.ac.uk}

\begin{abstract}
Topological insulators (TIs) are bulk insulators with exotic ‘topologically protected’ surface conducting modes. It has recently been pointed out that when stacked together, interactions between surface modes can induce diverse phases including the TI, Dirac semimetal, and Weyl semimetal. However, currently a full experimental understanding of the conditions under which topological modes interact is lacking. Here, working with multilayers of the TI Sb$_2$Te$_3$ and the band insulator GeTe, we provide experimental evidence of a multiple topological modes in a single Sb$_2$Te$_3$-GeTe-Sb$_2$Te$_3$ structure. Furthermore, we show that reducing the thickness of the GeTe layer induces a phase transition from a Dirac-like phase to a gapped phase. By comparing different multilayer structures we demonstrate that this transition occurs due to the hybridisation of states associated with different TI films. Our results demonstrate that the Sb$_2$Te$_3$-GeTe system offers strong potential towards manipulating topological states as well as towards controlledly inducing various topological phases.
\end{abstract}
\begin{document}

\flushbottom
\maketitle

\thispagestyle{empty}

\section*{Topological surface modes}

Topological insulators (TIs) are a recently emerged class of materials which are insulating in the bulk, but whose surface harbours `topologically protected' conducting modes~\cite{HasanKaneRMP2010}. The existence of the topological surface modes (TSMs) stems ultimately from the high spin-orbit coupling in TIs which `inverts' the conduction and valence bands. In other words, the band structure of the TI is topologically distinct from the ordinary band insulator (BI) and the TSMs arise because the band gap \textit{must} close at a TI-BI interface, or even an interface between a TI and the vacuum~\cite{MooreNature2010}. Importantly, local perturbations (e.g., disorder) which do not alter the topological properties of the system cannot localise the TSMs, and this gives rise to their `topological protection'. The TSMs have various exotic properties such as a linear \textit{Dirac}-like dispersion, and a well-defined helicity, i.e., spin and momentum vectors at fixed angles to each other. These properties render them useful in a variety of settings such as low-power electronics and spin-based communication and computation. In addition, when TSMs couple to each other they are predicted to produce a very diverse phase diagram of novel topological phases~\cite{BurkovBalentsPRL2011,Li_etal_SRep2014,Owerre_arXiv_2016}. In particular, in Ref.~{\renewcommand{\citemid}{}\cite[]{BurkovBalentsPRL2011}} it was first proposed that superlattices of alternating TI and BI layers can overall either be TIs or BIs depending on the specifics of how the TSMs couple, and under certain conditions be Weyl semimetals~\cite{HasanXuBian_PhysScripta2015}. Experimentally, there have been reports of TIs in superlattice structures including PbSe-Bi$_2$Se$_3$ structures~\cite{Nakayama_etal_PRL2012,SasakiSegawaAndoPRBR2014}, Bi$_14$Rh$_3$I$_9$~\cite{Rasche_etal_NatMater2013}, and Sb-Te binary systems~\cite{Takagaki_etal_JPCM2013,Takagaki_etal_JPCM2014}, although the crucial aspect of how the TSMs couple across the intervening layers remains unexplored. In this manuscript we investigate the low-temperature (low-$T$) electrical properties of molecular beam epitaxy (MBE)-grown bi-layer and tri-layer structures of Sb$_2$Te$_3$, a well-known TI~\cite{Zhang_etal_NPhys2009}, and GeTe, a narrow band gap semiconductor that goes superconducting at very low $T$~\cite{Hein_etal_PRL1964,Narayan_etal_PSSRRL2016}. Intriguingly, our results indicate two-dimensional (2D) transport, indicating that the structures are dominated by the Sb$_2$Te$_3$-GeTe interface. Remarkably, we realise specific situations in which the tri-layer system has either a \textit{quadratically dispersive} mode or \textit{three linearly dispersive} Dirac-like modes. Based on very simple assumptions we argue that the two states are topologically distinct, thus suggesting Sb$_2$Te$_3$-GeTe heterostructures to be promising platforms to induce topological phase transitions and also realise multi-TSM systems~\cite{BurkovBalentsPRL2011,Li_etal_SRep2014,Tominaga_etal_AdvMaterInt2014}.

TSMs in TIs have been most clearly identified using angle-resolved-photoemission-spectroscopy~\cite{Hsieh_etal_Nature2008,Hsieh_etal_Science2009}, although this technique is unable to probe the interior of materials, and thus has limited applicability in studying buried TSMs in structures of the type we report. Electrically, TSMs can be identified by observing quantum corrections to the conductivity $\sigma$ at low $T$. The strong spin-orbit interaction in TIs engenders positive quantum corrections to $\sigma$ in the form of weak antilocalization (WAL) at low $T$~\cite{HLN_ProgTheorPhys1980} which manifests as a characteristic cusp in the magnetoresistance described by the Hikami-Larkin-Nagaoka (HLN) equation~\cite{HLN_ProgTheorPhys1980}:

\begin{equation}
\label{HLN}
\Delta \sigma_{xx}^{2D} \equiv \sigma_{xx}^{2D}(B_{\perp}) - \sigma_{xx}^{2D}(0) = \alpha \frac{e^2}{2 \pi^2 \hbar}\left[ \ln \left( \frac{\hbar}{4eB_{\perp}\ell_{\varphi}^2}\right) - \Psi \left( \frac{1}{2} + \frac{\hbar}{4eB_{\perp}\ell_{\varphi}^2}\right) \right]
\end{equation}

Here $\sigma_{xx}$ indicates the longitudinal component of conductivity and the superscript 2D indicates that the equation is valid for a two-dimensional conducting sheet, $B_{\perp}$ is the magnetic field perpendicular to the 2D plane, $\alpha$ is a parameter = 0.5 for each 2D WAL channel, $e$ is the electronic charge, $\hbar$ is Planck's constant divided by 2$\pi$, $\ell_{\varphi}$ is the phase coherence length, and $\Psi$ is the digamma function. In TI thin films it is expected that $\alpha = 1$ due to the top and bottom surfaces, although often transport measurements in TI thin films are consistent with $\alpha \approx 0.5$, i.e., the existence of a single TSM~\cite{Veldhorst_etal_PSSRRL2013}. There is a growing consensus that this is due to the interfacial disorder between the TI and substrate which destroys the bottom TSM. Importantly, this hinders progress towards understanding the interaction between TSMs. As an alternative, exfoliated TI structures have been used~\cite{Checkelsky_etal_PRL2011} which are significantly less prone to interfacial disorder, but offer considerably less control over the film thickness. In this context, therefore, our approach of utilising MBE-grown multilayers is particularly promising in that it potentially overcomes both these limitations.

\begin{figure}
\centering
\includegraphics[width=\linewidth]{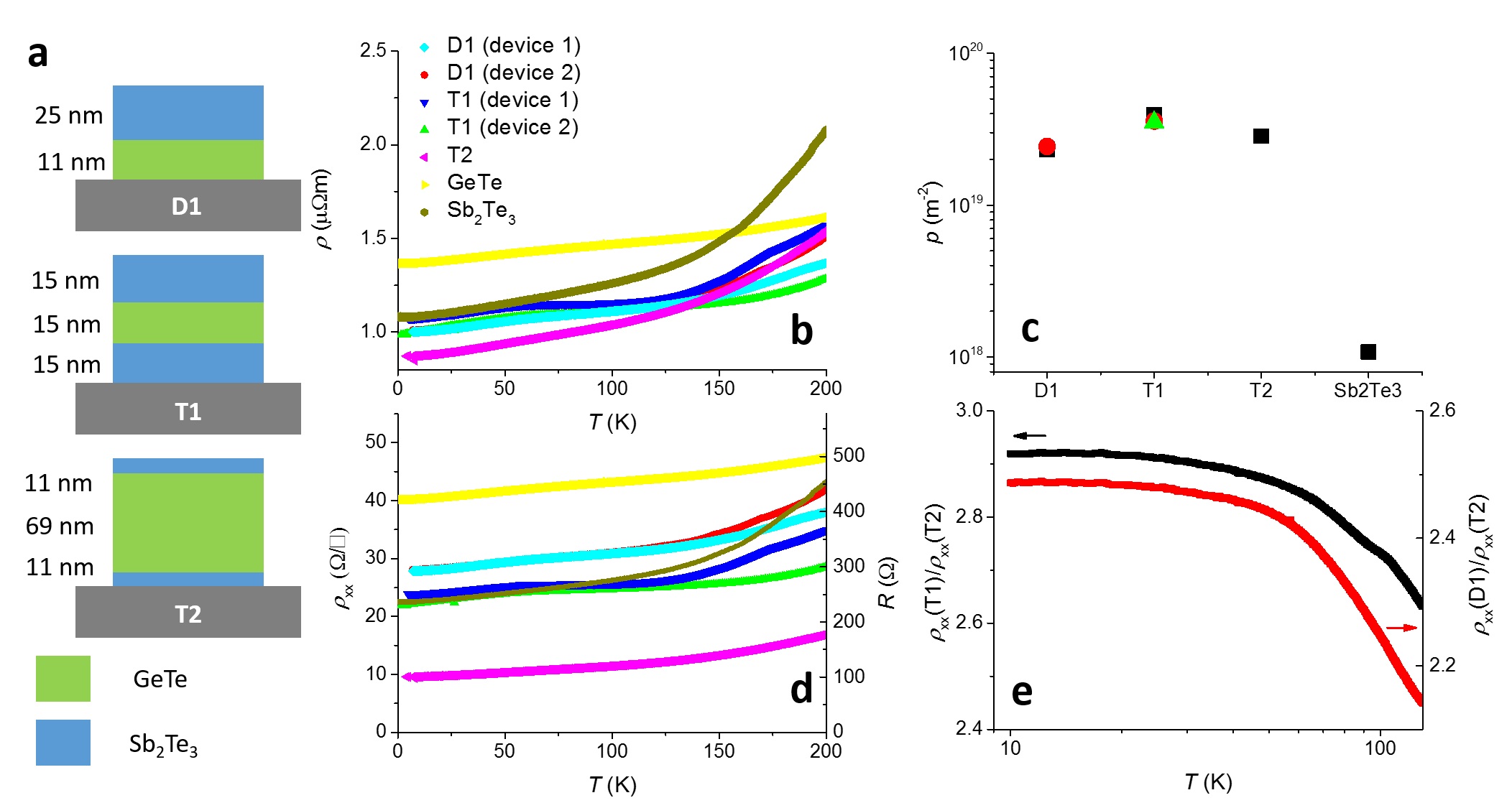}
\caption{The multilayer structures that are investigated are schematically depicted in (a). In (b) we show the $T$-dependence of $\rho$ where, intriguingly, it is seen that the heterostructures have a lower $\rho$ than both of the parent materials. (c) The carrier concentrations in the multilayers are more than an order of magnitude greater than that of pure Sb$_2$Te$_3$ and this is a direct outcome of the proximity to the larger bandgap material GeTe (measurements from multiple samples are shown for D1 and T1). In (d) the data shown in (b) is plotted as a 2D resistivity $\rho_{xx}$. The right axis shows the measured resistance $R$. (e) shows the ratio of $\rho_{xx}$ in T1 and T2 to be $\approx 2.9$, and that of D1 and T2 to be $\approx 2.5$ at low $T$.}
\label{fig1}
\end{figure}

\section*{Results and Discussion}

The wafer structures that we report here are schematically represented in Figure~\ref{fig1}a. The first is a double-layer structure with 11~nm of GeTe capped by 25~nm of Sb$_2$Te$_3$, henceforth referred to as D1. The second and third are three-layer Sb$_2$Te$_3$-GeTe-Sb$_2$Te$_3$ sandwich structures which we refer to as T1 and T2, respectively. In T1 each of the three layers is 15~nm thick, and T2 consists of a relatively thick 69~nm GeTe layer between 11~nm Sb$_2$Te$_3$ layers. The lattice mismatch of $\approx 2.5$\,\% between Sb$_2$Te$_3$ and GeTe is small enough to allow for coherent growth (see Methods). In addition, to benchmark our results, we have studied pure Sb$_2$Te$_3$ and GeTe films of thickness 48~nm and 34~nm, respectively. Figure~\ref{fig1}b shows the resistivity $\rho$ vs $T$ of Hall bar devices fabricated from the different wafers. The first important fact to note is that, surprisingly, the multilayer structures are \textit{consistently less resistive} than the pure samples, and closer in resistivity to Sb$_2$Te$_3$. Especially noteworthy in this context is that T2, which has the largest relative proportion of GeTe ($\approx 76$\,\%), has the lowest $\rho$ and shows the strongest deviation from pure GeTe. These observations clearly indicate that the individual layers do not behave simply as independent resistors connected in parallel, but rather interact with each other. One possibility is that the band bending at the interface between GeTe (band gap $\approx 0.65$~eV~\cite{Sante_etalAdvMater2013}) and Sb$_2$Te$_3$ (band gap $\approx 0.2$~eV~\cite{MadelungRosslerSchulz_Sb2Te3book_1998}) serves to draw and confine additional charge carriers into Sb$_2$Te$_3$, and thus enhance its conductivity. This notion is corroborated by the measured areal carrier densities in D1, T1 and T2 being over an order of magnitude greater than in Sb$_2$Te$_3$ (Figure~\ref{fig1}c). Here we consider the areal rather than bulk carrier density since at low $T$ the conduction in Sb$_2$Te$_3$ is largely through the TSMs and thus expected to be two-dimensional (2D)~\cite{Zhang_etal_NPhys2009}. As a reference, the bulk carrier concentration in GeTe was measured to be $\approx 5.5 \times 10^{26}$~m$^{-3}$ which, when scaled by the thickness of GeTe is $\sim 10^{19}$~m$^{-2}$, is comparable with D1, T1 and T2. In Figure~\ref{fig1}d we plot the 2D resistivity $\rho_{xx} \equiv \rho \times t$, where $t$ is the thickness of the film, as a function of $T$: in order of increasing $\rho_{xx}$ we find T2, followed by T1 and Sb$_2$Te$_3$ which are very similar below $\approx 100$~K, followed by D1 and then GeTe. The strong dissimilarity between the characteristics of T1 and T2 suggests that there is more physics underlying the heterostructures than just band bending. We return to this fact after having inspected the low T electrical characteristics.

\begin{figure}[ht]
\centering
\includegraphics[width=\linewidth]{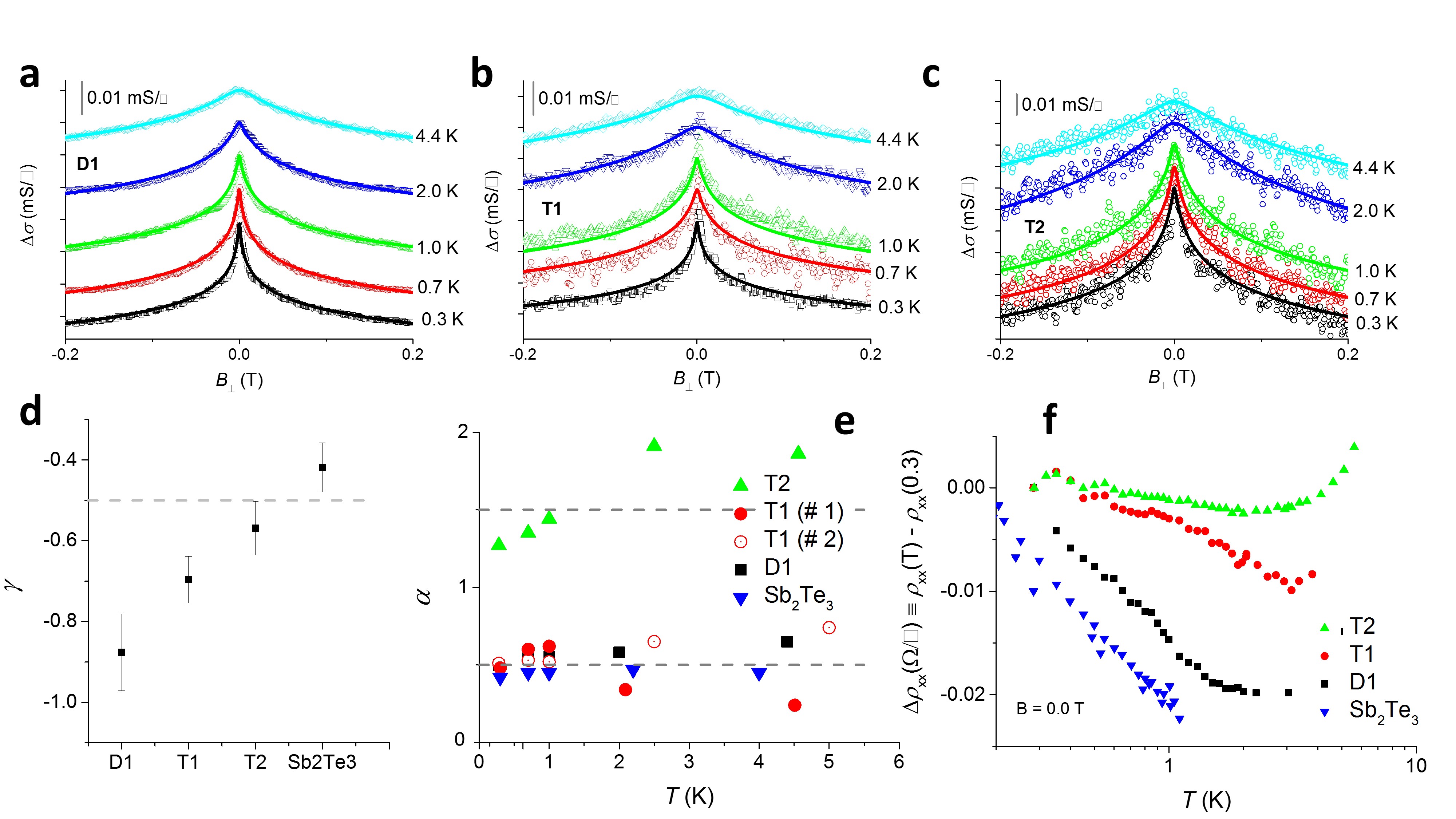}
\caption{(a) – (c) show the WAL characteristics for D1, T1 and T2, respectively. Also shown are fits to the HLN formula (Equation 1). The traces are offset vertically for clarity. In (d) we observe that the rate of decoherence in T2 and Sb$_2$Te$_3$ is consistent with that expected due to inter-particle interactions, whereas in D1 and T1 the rate of decoherence is higher. (e) The parameter $\alpha$ which is a measure of the number of WAL channels contribute to the transport is seen to be 0.5 (corresponding to 1 WAL channel) for all the wafers except T2 in which $\alpha$ is approximately 1.5, suggesting the presence of three 2D WAL channels. (f) At low-$T$ all the wafers show a logarithmic increase in $\rho_{xx}$ as is expected in 2D systems. It is conceivable that in T1 and T2 GeTe is showing hints of superconductivity at $\approx 0.35$~K, but this is unlikely to influence the transport unless $T$ is reduced significantly.}
\label{fig2}
\end{figure}

In Figure~\ref{fig2}a, b and c we show the low-field magnetoconductivity $\sigma_{xx} \equiv \rho_{xx}/(\rho_{xx}^2 + \rho_{xy}^2)$ of D1, T1 and T2, respectively, and observe a pronounced cusp-like minimum around $B_{\perp} = 0$~T. Here $\rho_{xy}$ is the Hall resistivity. The signal, expectedly, is not consistent with bulk WAL (see supplementary Figure S1) where $\Delta \sigma_{xx} \equiv \sigma_{xx}(B_{\perp}) - \sigma_{xx}(0) \sim B_{\perp}^{1/2}$~\cite{Kawabata_JPSP1980}, but instead is very well-described by the HLN formula (Equation~\ref{HLN}) with $\alpha$ and $\ell_{\varphi}$ as fit parameters. We find that $\ell_{\varphi}$ decays as a power law $\sim T^{-\gamma}$ where the exponent $\gamma$ is plotted in Figure~\ref{fig2}d. Apart from Sb$_2$Te$_3$ and T2, all the structures are observed to undergo decoherence significantly faster than the $\gamma = 0.5$ expected in 2D Nyquist scattering due to inter-particle interactions~\cite{AltshulerTagliacozzoTognetti_book_2003}. Figure~\ref{fig2}e shows that $\alpha$ is $\approx 0.5$ for Sb$_2$Te$_3$, T1 and D1, but is $\approx$~1.5 for T2 (see supplementary Figure S2 for error estimates on $\alpha$). Recalling that $\alpha$ gets a contribution of 0.5 for each 2D WAL state, we conclude that Sb$_2$Te$_3$, T1 and D1 have one 2D mode each while T2 has \textit{three}. While this is consistent with T2 being the least resistive sample under investigation (Figure~\ref{fig1}), there are two important points to consider: First, as discussed earlier in the manuscript, it is expected that $\alpha \approx 0.5$ for Sb$_2$Te$_3$ since the TSM at the substrate interface is destroyed. However, unexpectedly we find that even in D1, in which the Sb$_2$Te$_3$ film is grown on a well-lattice-matched MBE-grown GeTe film, a second TSM is \textit{not} observed. Second, the occurrence of three 2D modes in T2 agrees with the three TI-BI interfaces (apart from the TI-substrate interface), but this is in apparent contradiction with the observation of only one 2D mode in T1. In a previous study it was found that bare GeTe thin films with $t = 34$~nm have $\alpha \approx 1$~\cite{Narayan_etal_PSSRRL2016}, i.e., one WAL mode on each side of the film, but GeTe has been intentionally omitted from this plot since it shows superconducting correlations below $T \approx 1.5$~K which strongly influence the HLN fits~\cite{Narayan_etal_PSSRRL2016}.

In the following we investigate the precise role played by GeTe in these heterostructures. First, in order to check whether the tendency of GeTe to become superconducting at very low $T$ has any influence on the transport properties of the multilayer structures we inspect the $T$-dependence of $\rho_{xx}$ for $T < 10$~K in Figure~\ref{fig2}f. We find clearly that $\mbox{d}\rho_{xx}/\mbox{d}T$ is negative below $\approx 3$~K, qualitatively similar to bare Sb$_2$Te$_3$ and in strong agreement with our hypothesis that transport in the multilayers is largely confined to Sb$_2$Te$_3$. There may be a maximum in T1 and T2 around 0.35~K, but even if this does signal the onset of superconductivity, it is unlikely that this perceptibly affects the transport. Indeed, within the picture that the band gaps of Sb$_2$Te$_3$ and GeTe conspire to inject charges from the latter to the former, it is evident why superconductivity is not observed in these samples: the superconducting temperature $T_c$ of GeTe is a monotonically increasing function of its carrier concentration~\cite{Hein_etal_PRL1964} and thus reducing the chemical potential in GeTe, as is achieved when placing it in close proximity to Sb$_2$Te$_3$, will only further suppress $T_c$. At the typical carrier concentrations obtained in the bare GeTe samples $T_c \approx 0.1$~K~\cite{Narayan_etal_PSSRRL2016}, and this essentially precludes the possibility of observing superconductivity at the temperatures our experiments are performed at ($\geq 0.3$~K). Yet, the strong variation of $\rho_{xx}$ between the different multilayer structures clearly indicates that GeTe is not playing a passive role in the transport. This is especially seen in the contrast between the tri-layer structures T1 and T2: it is seen that $\rho_{xx}$(T1) $\approx$ 3$\rho_{xx}$(T2), corresponding almost exactly with the ratio of the number of 2D modes as given by the HLN formula (Figure~\ref{fig1}d). Since the data are clearly indicative of missing WAL modes in T1, we conjecture that certain TSMs in the measured heterostructures hybridise~\cite{Zhang_etal_NPhys2010,Murakami_Hybridisationbook_2015} and become \textit{gapped}. Importantly, this mechanism would undermine their topological nature and modify the 2D band dispersion from being linear, Dirac-like, to being parabolic near the band minimum.

\begin{figure}[ht]
\centering
\includegraphics[width=\linewidth]{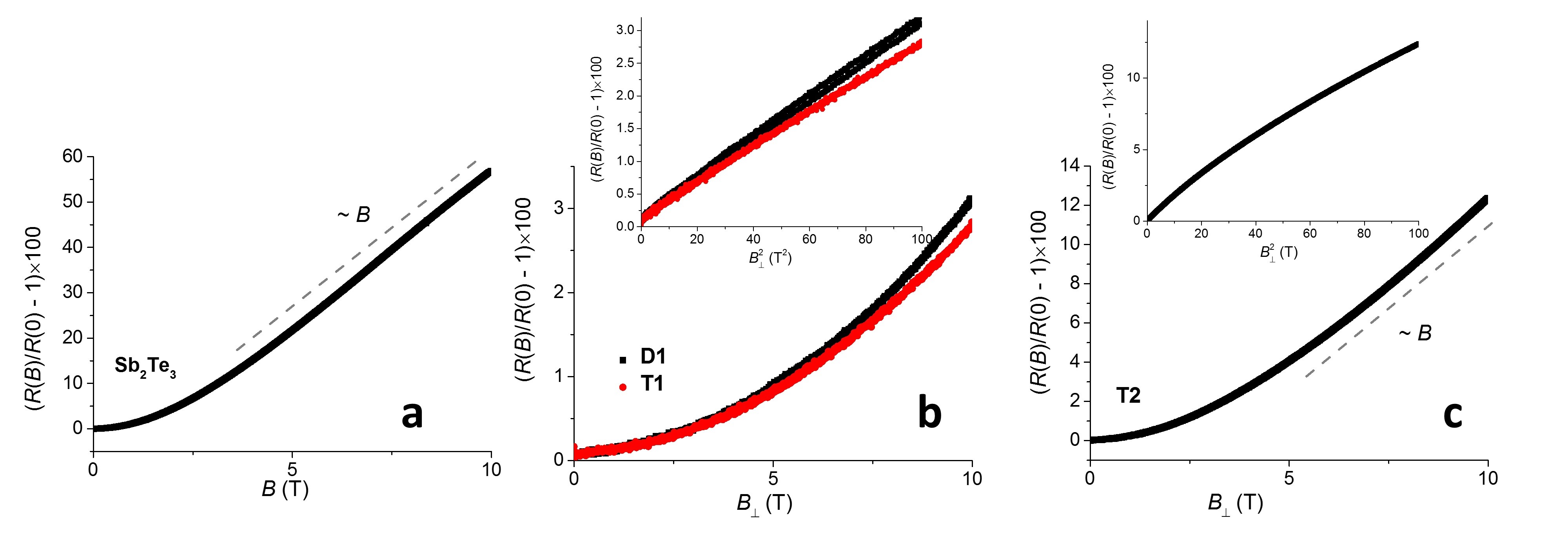}
\caption{The high-field magnetoresistance of Sb$_2$Te$_3$ (a) and T2 (c) is qualitatively different from D1 and T1 (b), with the latter being quadratic in $B_{\perp}$ over the entire field range explored. Moreover, it is observed in (b) that the magnetoresistance characteristics of D1 and T1 are almost identical, thus suggesting similar underlying physics. The insets in (b) and (c) show the magnetoresistance as a function of $B_{\perp}^2$, clearly bringing out the contrasting behaviours of T1/D1 when compared to T2. The observed linear magnetoresistance in Sb$_2$Te$_3$ and T2 is a commonly observed feature in materials with a linear dispersion relation.}
\label{fig3}
\end{figure}

To investigate the proposed scenario further we study the high-field magnetoresistance characteristics. Figure~\ref{fig3}a exemplifies a commonly observed feature in TIs (or Dirac materials, to be more precise), namely a large and \textit{linear} magnetoresistance~\cite{Qu_etal_Science2010,Tang_etal_ACSNano2011,Wang_etal_PRL2012,Veldhorst_etal_PSSRRL2013}. There are various circumstances which can effect such behaviour including very large carrier concentrations~\cite{Veldhorst_etal_PSSRRL2013}, transport in `heavily' disordered conductors~\cite{ParishLittewood_Nature2003}, and the exclusive occupancy of the lowest Landau level~\cite{Abrikosov_JETP1969,AbrikosovPRB1998}. The first two of these are not applicable to our experimental system since the carrier concentration, which is similar between the multilayers, clearly does not influence the nature of magnetoresistance (Figure~\ref{fig3}), and the level of disorder in MBE-grown films is expected to be small. On the other hand, the third criterion is known to be experimentally more accessible in Dirac materials than those with a parabolic dispersion~\cite{Veldhorst_etal_PSSRRL2013}. That is, in our experimental system we can directly correlate the presence or absence of large, linear magnetoresistance to the presence or not of Dirac-like states. Thus, a critical conclusion we arrive at from Figure~\ref{fig3}a and \ref{fig3}b is that D1 is not, as one would naively expect, simply a TI layer on a lattice-matched, insulating substrate, but has at least one gapped state which dominates the magnetoresistance at high fields. Since the Sb$_2$Te$_3$ layer is significantly thicker than 6~nm, the approximate thickness below which the top and bottom TSMs are known to hybridise in Sb/Bi-based TIs~\cite{Zhang_etal_NPhys2010,Li_etal_AdvMater2010,Jiang_etal_PRL2012} it is unlikely that the top TSM has developed a gap. Which then implicates the lower TSM, but raises the question as to what it hybridises with. Recalling that 34~nm thick GeTe harbours \textit{two} surface states~\cite{Narayan_etal_PSSRRL2016}, our data strongly points in favour of the 2D modes on either side of the 11~nm thick GeTe layer hybridising with each other. This is also consistent with the fact that T1 has only one 2D WAL mode where, as per the previous arguments, the two embedded TSMs hybridise across the 15~nm thick GeTe layer. This idea is firmly reinforced by the data in Figure~\ref{fig3}b where we find that the magnetoresistance of D1 and T1 agree \textit{quantitatively}, thus indicating similar underlying physics. Of particular interest in this regard is the fact that in T1, both hybridising states are topological in nature, but in D1, the bottom state is necessarily not topological (but Dirac-like~\cite{Krempasky_etal}). The hybridisation picture also provides an estimate of the wavefunction extent outward from the interface. Since hybridisation is possible across a 15~nm thick GeTe layer, we can conclude that the wavefunctions on either side have significant overlap and therefore must have a spatial extent of $\geq 7.5$~nm. Clearly, however, the wavefunction is not symmetric about the GeTe-Sb$_2$Te$_3$ interface since its extent in Sb$_2$Te$_3$ is $\leq 5.5$~nm. Such asymmetry is not unexpected being dependent on factors such as the band offsets between the different materials and precise form of the interface confining potential.	
%

The conclusion that the buried TSMs in T1 hybridise across the GeTe layer can also be arrived at by contrasting the behaviours of T1 and T2: Figure~\ref{fig2}e suggest that none of the TSMs in T2 hybridise and this is buttressed by the observation in Figure~\ref{fig3}c that  $\rho_{xx}$ vs $B_{\perp}$ is linear above $\approx 5$~T. Therefore, since the TSMs flanking the Sb$_2$Te$_3$ layer in T2 do not hybridise, they must not in T1 either where the Sb$_2$Te$_3$ layers are thicker. Thus the only possibility is that we are observing the hybridisation of TSMs across the GeTe layer, i.e., between TSMs associated with \textit{different} TI layers. To the best of our knowledge this is the first experimental demonstration of hybridisation between TSMs not related by symmetry. More importantly, our data evidences a critical thickness of GeTe at which the Sb$_2$Te$_3$-GeTe-Sb$_2$Te$_3$ heterostructure undergoes a topological phase transition, thus indicating it to be a promising system in which to study topological phase transitions~\cite{Kim_etal_PRBR2010,Sa_etal_PRL2012,Kim_etal_PRL2012,Tominaga_etal_AdvMaterInt2014}.

But then what of the remaining TSM on the top surface of D1 and T1? Being gapless, this mode will dominate the low $T$ transport characteristics, but the high-field transport will be dominated by the non-Dirac-like modes which have a dramatically lower magnetoresistance. Notably, it was shown in Ref.~{\renewcommand{\citemid}{}\cite[]{LuShiShen_PRL_2011}}. that the size of the WAL effect should reduce as the gap in the surface states widens relative to the Fermi energy, i.e, we do not expect the gapped states to contribute significantly to the measured WAL. In this context, Figure 1e where we find the ratios $\rho_{xx}$(T1)/$\rho_{xx}$(T2) and $\rho_{xx}$(D1)/$\rho_{xx}$(T2) to be be constant below $T \approx 30$~K, sheds light on the relevant energy scales. Both the ratios are $\sim 3$, consistent with the ratio of the number of 2D modes. This temperature scale corresponds to $\approx 2.6$~meV, which is two orders of magnitude smaller than the band gap of Sb$_2$Te$_3$, and thus the decrease in $\rho_{xx}$(T1)/$\rho_{xx}$(T2) and $\rho_{xx}$(D1)/$\rho_{xx}$(T2) with increasing $T$ most likely corresponds to the activation of the hybridised TSMs. We, therefore, have obtained a crude measure of the hybridisation gap to be about 2.6~meV. Not only does this provide an experimentally measured parameter useful for simulations of TI-BI superlattices, it also shows the experimental system to be a versatile platform to realise and manipulate multiple TSMs.
In conclusion,we have demonstrated a phase transition in the topological properties of Sb$_2$Te$_3$-GeTe-Sb$_2$Te$_3$ structures as the thickness of the GeTe layer is varied. We have demonstrated this system to be a close realisation of the model proposed by Burkov and Balents~\cite{BurkovBalentsPRL2011} to induce novel topological phases. We have provided the first experimental demonstration of hybridisation between TSMs associated with different TIs and shown the experimental system to be potentially exploitable in applications requiring multiple TSMs in parallel.

\section*{Experimental Section}

\subsection*{Thin film growth} The GeTe and Sb$_2$Te$_3$ films were grown on Si(111) wafers using MBE. Before the deposition, the Si substrates were cleaned by the HF-last RCA procedure to remove the native oxide and passivate the surface with hydrogen.
Subsequently, the substrates were heated in-situ at 750~$^{\circ}$C for 20 min to desorb hydrogen atoms from the surface. The Ge, Sb, and Te material fluxes were generated by effusion cells with temperatures of 1250~$^{\circ}$C, 440~$^{\circ}$C, and 330~$^{\circ}$C, respectively. For all samples the Te shutter was opened 2 seconds before the Ge or Sb shutter in order to saturate the Si substrate surface with Te. During the entire growth process the substrate was set at 280~$^{\circ}$C. The lattice parameters of Sb$_2$Te$_3$ and GeTe are, respectively, a$_{\mathrm{Sb2Te3}} = 4.26$~{\AA} and a$_{\mathrm{GeTe}} = 8.32$~{\AA}, and thus the lattice mismatch at 2:1 correspondence is 
(2 $\times$ a$_{\mathrm{Sb2Te3}}$ – a$_{\mathrm{GeTe}}$)/a$_{\mathrm{GeTe}}$ = 2.5\,\%.
\subsection*{Device fabrication and electrical measurements} We used photolithography and argon ion milling to fabricate Hall bar devices of dimensions 100~$\mu$m~$\times$~1050~$\mu$m, after which Ti/Au ohmic contacts were deposited using a lift-off process. The devices were packaged and measured in a He 3 cryostat with a base $T$ = 280~mK, and equipped with a 10~T superconducting magnet. Resistance and Hall measurements were made in a standard four-terminal setup with an excitation current $I_{ex} = 0.1$ – $1~\mu$A at frequency $f$ = 17 Hz.

\bibliography{References}

\section*{Acknowledgements} T.-A.N., D.B., D.R., and V.N. acknowledge funding from the Leverhulme Trust, UK, T.-A.N., D.B., A.S., R.M., C.B., D.R., and V.N. acknowledge funding from EPSRC (UK). G.M., M.L. and D.G. acknowledge financial support from the DFG-funded priority programme SPP1666. V.N. acknowledges useful discussions with Michael Pepper.

\section*{Author contributions statement}
T.-A.N., D.B., and V.N. performed the experiments, T.-A.N., D.B., R.M. and A.S. fabricated the devices, G.M., M.L., and D.G. grew the samples, V.N. wrote the paper with inputs from R.M., G.M., D.B., and T.-A.N. All the authors reviewed the manuscript.

\section*{Additional information}

\textbf{Competing financial interests} The authors declare no competing financial interests.

\setcounter{figure}{0} \renewcommand{\figurename}{Fig. S}

\newpage

\begin{center}

\large \textbf{Topological states and phase transitions in Sb$_2$Te$_3$-GeTe multilayers - Supplementary Material}

\end{center}

\begin{figure*}[ht]
\centering
\includegraphics[width=6.5in]{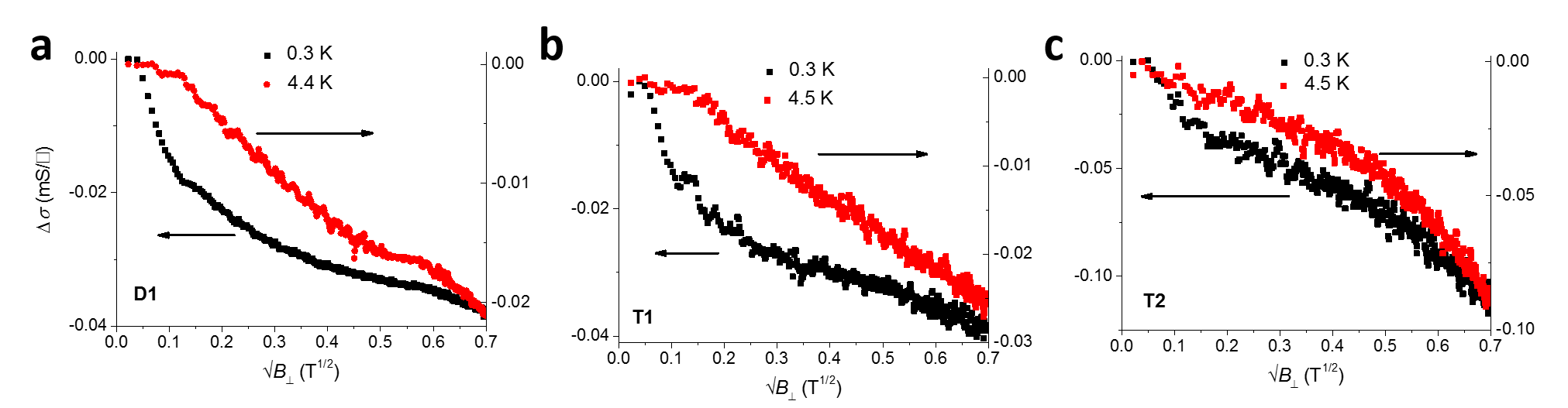}
\caption{(a) – (c) show the WAL corrections as a function of $\sqrt{B_{\perp}}$. Clearly, the data are not consistent with bulk WAL where $\Delta \sigma_{xx} \sim B_{\perp}^{1/2}$~\cite{Kawabata_JPSP1980}. The black and red traces correspond to $\Delta \sigma_{xx}$ measured at 0.3~K and 4.5~K, respectively.}
\label{SM_fig1}

\includegraphics[width=5.75in]{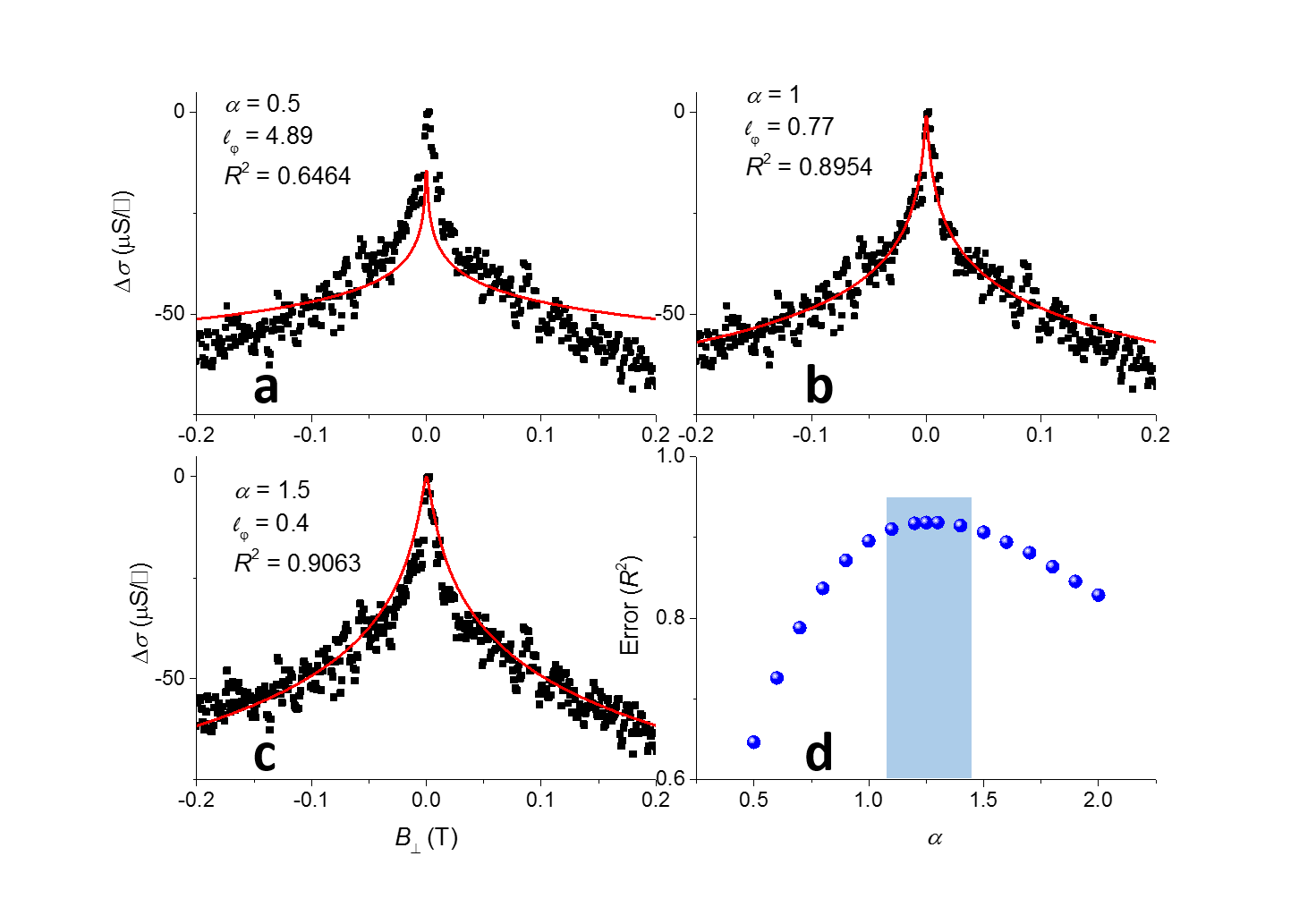}
\caption{(a) – (c) compare the experimental data (symbols) and HLN fit (solid line) for the fitting parameters listed in each panel. Clearly, $1 < \alpha < 1.5$ yield the best fits, consistent with three TSMs. In order to produce these graphs, $\alpha$ was held constant and $\ell_{\varphi}$ varied to obtain the best fit. The $R^2$ measure of the fit, i.e., the squared sum of difference between the experimental and theoretical curves was used to gauge the quality of the fit. As is shown in (d), the $R^2$ measure shows a maximum in the range of $\alpha = 1.25$.}
\label{SM_fig2}
\end{figure*}

\end{document}